\documentclass{elsart}
 \usepackage{latexsym}
\usepackage{amssymb}

\begin{document}
\begin{frontmatter}

\title{Influence of the beam-size or MD-effect on particle losses at
B-factories\\ PEP-II and KEKB}

\author{G.L.~Kotkin\thanksref{RFBR}, V.G.~Serbo\thanksref{ead}}
\thanks[RFBR]{This work is supported in part by INTAS
(code 00-00679), RFBR (code 02-02-17884) and by University of
Russia (code 02.01.005)}
 \thanks[ead]{Corresponding author. E-mail: serbo@math.nsc.ru}

\address{Novosibirsk State University, 630090 Novosibirsk, Russia}

\begin{abstract}
For the $e^+ e^- \rightarrow e^+ e^- \gamma$ process at colliding
beams, macroscopically large impact parameters give an essential
contribution to the standard cross section. These impact
parameters may be much larger than the transverse sizes of the
colliding bunches. It means that the standard calculations have to
be essentially modify. In the present paper such a beam-size or
MD-effect is calculated for bremsstrahlung at B-factories PEP-II
and KEKB using the list of nominal parameters from Review of
Particle Physics (2002). We find out that this effect reduces beam
losses due to bremsstrahlung by about 20\%.
\end{abstract}
\end{frontmatter}

\section{Introduction: beam-size or MD-effect}

The so called beam-size or MD-effect is a phenomenon discovered in
experiments \cite{Blinov82} at the MD-1 detector (the VEPP-4
accelerator with $e^+e^-$ colliding beams , Novosibirsk 1981). It
was found out that for ordinary bremsstrahlung, macroscopically
large impact parameters should be taken into consideration. These
impact parameters may be much larger than the transverse sizes of
the interacting particle bunches. In that case, the standard
calculations, which do not take into account this fact, will give
incorrect results. The detailed description of the MD-effect can
be found in review \cite{KSS}.

We start with a few words about history of this effect. In
$1980$--$1981$ a dedicated study of the process $e^+ e^-
\rightarrow e^+ e^- \gamma$ has been performed at the collider
VEPP-$4$ in Novosibirsk using the detector MD-1 for an energy of
the electron and positron beams $E_e=E_p = 1.8$ GeV and in a wide
interval of the photon energy $E_\gamma$ from $0.5$ MeV to
$E_\gamma \approx E_e$. It was observed \cite{Blinov82} that the
number of measured photons was smaller than that expected. The
deviation from the standard calculation reached $30 \%$ in the
region of small photon energies and vanished for large energies of
the photons. A.~Tikhonov \cite{Tikhonov82} pointed out that those
impact parameters $\varrho$, which give an essential contribution
to the standard cross section, reach values of $\varrho_m \sim 5$
cm whereas the transverse size of the bunch is $\sigma_\perp \sim
10^{-3}$ cm. The limitation of the impact parameters to values
$\varrho \lesssim \sigma_\perp$ is just the reason for the
decreasing number of observed photons.

The first calculations of this effect have been performed in Refs.
\cite{BKS} and \cite{BD} using different versions of
quasi--classical calculations in the region of large impact
parameters. Further experiments, including the measurement of the
radiation probability as function of the beam parameters,
supported the concept that the effect arises from the limitation
of the impact parameters. Later on, the effect of limited impact
parameters was taken into account when the single bremsstrahlung
was used for measuring the luminosity at the VEPP--$4$
collider~\cite{Blinov88} and at the LEP-I collider~\cite{Bini94}.
In the case of the VEPP--$4$ experiment~\cite{Blinov88}, it was
checked that the luminosities obtained using either this process
or using other reactions (such as the double bremsstrahlung
process $e^+e^- \rightarrow e^+ e^- \gamma \gamma$, where the
MD-effect is absent) agreed with each other.

A general scheme  to calculate the finite beam size effect had
been developed in paper~\cite{KPS85a} starting from the quantum
description of collisions as an interaction of wave packets
forming bunches. Since the effect under discussion is dominated by
small momentum transfer, the general formulae can be considerably
simplified. The corresponding approximate formulae were given.
They are obtained from an analysis of Feynman diagrams and it
allows to estimate the accuracy of approximation. In a second
step, the transverse motion of the particles in the beams can be
neglected. The less exact (but simpler) formulae, which are then
found, correspond to the results of Refs.~\cite{BKS} and
\cite{BD}. It has also been shown that similar effects have to be
expected for several other reactions such as bremsstrahlung for
colliding $ep$--beams~\cite{KPS85b}, \cite{KPSS88}, $e^+e^-$--
pair production in $e^\pm e$ and $\gamma e$
collisions~\cite{KPS85a}. The corresponding corrections to the
standard formulae are now included in programs for simulation of
events at linear colliders. The influence of MD-effect on
polarization had been considered in Ref.~\cite{KKSS89}.

In 1995 the MD-effect was experimentally  observed at the
electron-proton collider HERA~\cite{Piot95} on the level predicted
in~\cite{KPSS88}.

The possibility to create high-energy colliding $\mu^+\mu^-$ beams
is now wildly discussed. For several processes at such colliders a
new type of beam-size effect will take place --- the so called
linear beam-size effect~\cite{KMS}. The calculation of this effect
had been performed by method developed for MD-effect
in~\cite{KPS85a}.

It was realized in last years that  MD-effect in bremsstrahlung
plays important role for the problem of beam lifetime. At storage
rings TRISTAN and LEP-I, the process of a single bremsstrahlung
was the dominant mechanism for the particle losses in beams. If
electron loses more than $1\;\%$ of its energy, it leaves the
beam. Since  MD-effect reduced considerable the effective cross
section of this process, the calculated beam lifetime in these
storage rings was larger by about $25 \; \%$ for
TRISTAN~\cite{Funakoshi} and by about $40 \; \%$ for
LEP-I~\cite{Burkhard} (in accordance with the experimental data)
then without taken into account the MD-effect.

In next Section we present the qualitative description of the
MD-effect. In Sect. 3 we calculate the MD-effect and its influence
on the beam losses at the existing B-factories. We find out that
this effect reduces beam losses due to bremsstrahlung by about
20\%.

At the end of this section we also mention about recent
paper~\cite{BK02} in which previous results~\cite{BKS}, \cite{BD},
\cite{KPS85a} about bremsstrahlung spectrum had been revised. It
was claimed that an additional subtraction related to the coherent
contribution has to be done. However, this additional subtraction
is negligible small for the real parameters of the existing
B-factories. Besides, paper~\cite{BK02}, in our opinion, is
incorrect. In our critical remark~\cite{KS02} we analyzed in
detail the coherent and incoherent contributions in the
conditions, considered in paper~\cite{BK02}, and, in contrast to
above claims, we found out that under these conditions the
coherent contribution is completely negligible and, therefore,
there is no need to revise the previous results.

\section{Qualitative description of the MD-effect}

\begin{figure}[!htb]
  \centering
  \setlength{\unitlength}{1cm}
\unitlength=2.0mm \special{em:linewidth 0.4pt}
\linethickness{0.4pt}
\begin{picture}(26.00,15.00)

\put(1.00,1.80){\line(1,0){24.00}}
\put(1.00,1.60){\line(1,0){24.00}}
\put(25.2,1.70){\vector(1,0){0.10}}

\put(1.00,9.20){\vector(1,0){6.00}}
\put(14.00,10.00){\circle{3.40}} \put(7.00,9.2){\line(1,0){5.3}}
\put(15.70,9.20){\vector(1,0){10.00}}
\put(15.0,11.5){\line(1,0){0.8}} \put(17,11.5){\line(1,0){0.8}}
\put(19,11.5){\line(1,0){0.8}} \put(21,11.5){\line(1,0){0.8}}
\put(23,11.5){\vector(1,0){2.00}}

\put(14.00,4.50){\vector(0,1){2.00}}
\put(14.00,2.90){\line(0,1){0.71}}
\put(14.00,1.80){\line(0,1){0.51}}
\put(14.00,7.50){\line(0,1){0.7}}

\put(1.00,10.4){\makebox(0,0)[cc]{$E_e$}}
\put(1.00,3.00){\makebox(0,0)[cc]{$E_p$}}
\put(22.00,12.70){\makebox(0,0)[cc]{$E_\gamma$}}
\put(22.00,5.00){\makebox(0,0)[cc]{$\hbar q=(\hbar\omega/c,\,
\hbar{\bf q})$}}

\end{picture}
    \caption{Block diagram of radiation by the electron.}
 \label{fig:1}
  \end{figure}
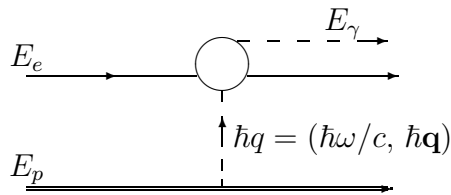

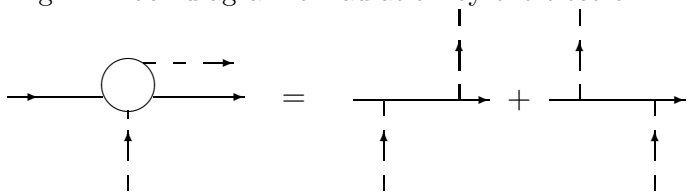
\begin{figure}[!htb]
  \centering
\unitlength=2.00mm \special{em:linewidth 0.4pt}
\linethickness{0.4pt}
\begin{picture}(47.00,15.00)
\put(2.00,9.20){\vector(1,0){2.00}}
\put(10.00,10.00){\circle{3.40}} \put(3.00,9.20){\line(1,0){5.30}}
\put(11.70,9.20){\vector(1,0){6.00}}
\put(10.00,5.00){\vector(0,1){2.00}}
\put(10.00,3.00){\line(0,1){1.00}}
\put(10.00,7.80){\line(0,1){0.50}}
\
\
\put(11.00,11.50){\line(1,0){0.80}}
\put(13.00,11.50){\line(1,0){0.80}}
\put(15.00,11.50){\vector(1,0){2.00}}
\
\put(25.00,9.00){\vector(1,0){9.00}}
\put(38.00,9.00){\vector(1,0){9.00}}

\put(27.00,8.00){\line(0,1){1.00}}
\put(27.00,5.00){\vector(0,1){2.00}}
\put(27.00,3.00){\line(0,1){1.00}}

\put(32.00,9.00){\line(0,1){1.00}}
\put(32.00,11.00){\vector(0,1){2.00}}
\put(32.00,14.00){\line(0,1){1.00}}

\put(40.00,9.00){\line(0,1){1.00}}
\put(40.00,11.00){\vector(0,1){2.00}}
\put(40.00,14.00){\line(0,1){1.00}}

\put(45.00,8.00){\line(0,1){1.00}}
\put(45.00,5.00){\vector(0,1){2.00}}
\put(45.00,3.00){\line(0,1){1.00}}

\put(21.00,9.00){\makebox(0,0)[cc]{=}}
\put(36.00,9.00){\makebox(0,0)[cc]{+}}

\end{picture}
 \caption{Compton scattering of equivalent photon on the electron.}
 \label{fig:2}
\end{figure}

Qualitatively we describe the MD--effect using as an example the
$e p \rightarrow e p \gamma$ process\footnote{Below we use the
following notations: $N_e$ and $N_p$ are the numbers of electrons
and protons (positrons) in the bunches, $\sigma_H$ and $\sigma_V$
are the horizontal and vertical transverse sizes of the proton
(positron) bunch, $\gamma_e=E_e/(m_ec^2)$, $\gamma_p=E_p/(m_pc^2)$
and $r_e=e^2/(m_e c^2)$ is the classical electron radius.}. This
reaction is defined by the diagrams of Fig.~\ref{fig:1} which
describe the radiation of the photon by the electron (the
contribution of the photon radiation by the proton can be
neglected). The large impact parameters $\varrho \gtrsim
\sigma_\perp$, where $\sigma_\perp$ is the transverse beam size,
correspond to small momentum transfer $\hbar q_\perp \sim (\hbar /
\varrho) \lesssim (\hbar / \sigma_\perp)$. In this region, the
given reaction can be represented as a Compton scattering
(Fig.~\ref{fig:2}) of the equivalent photon, radiated by the
proton, on the electron. The equivalent photons with frequency
$\omega$ form a ``disk'' of radius $\varrho_m \sim \gamma_p c /
\omega$ where $\gamma_p = E_p / (m_p c^2)$ is the Lorentz-factor
of the proton. Indeed, the electromagnetic field of the proton is
$\gamma_p$--times contracted in the direction of motion.
Therefore, at distance $\varrho$ from the axis of motion a
characteristic longitudinal length of a region occupied by the
field can be estimated as $\lambda \sim \varrho / \gamma_p$ which
leads to the frequency $\omega \sim c / \lambda \sim \gamma_p c /
\varrho$.

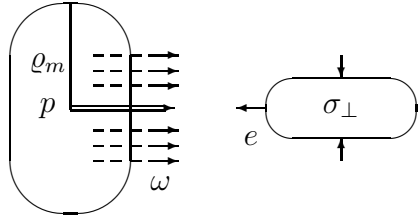
\begin{figure}[!htb]
  \centering
\unitlength=2.00mm \special{em:linewidth 0.4pt}
\linethickness{0.4pt}
\begin{picture}(47.00,15.00)

\put(13.00,7.00){\oval( 8,14)}

\put(13.00 ,6.85){\line(1,0){6.30}}
 \put(13.00,7.15){\line(1,0){6.30}}
 \put(13.00 ,6.90){\line(0,1){7.00}}
\put(19.40 ,7.00){\vector(1,0){0.50}}

\put(14.50 ,8.50){\line(1,0){0.70}}
 \put(15.70,8.50){\line(1,0){0.70}}
  \put(17.00,8.50){\line(1,0){0.70}}
 \put(18.20 ,8.50){\vector(1,0){2.00}}

\put(14.50 ,9.50){\line(1,0){0.70}}
\put(15.70,9.50){\line(1,0){0.70}}
 \put(17.00,9.50){\line(1,0){0.70}}
\put(18.20 ,9.50){\vector(1,0){2.00}}

\put(14.50 ,10.50){\line(1,0){0.70}}
 \put(15.70,10.50){\line(1,0){0.70}}
  \put(17.00,10.50){\line(1,0){0.70}}
\put(18.20 ,10.50){\vector(1,0){2.00}}

\put(14.50 ,5.50){\line(1,0){0.70}}
 \put(15.70,5.50){\line(1,0){0.70}}
  \put(17.00,5.50){\line(1,0){0.70}}
  \put(18.20 ,5.50){\vector(1,0){2.00}}

\put(14.50 ,4.50){\line(1,0){0.70}}
 \put(15.70,4.50){\line(1,0){0.70}}
  \put(17.00,4.50){\line(1,0){0.70}}
 \put(18.20 ,4.50){\vector(1,0){2.00}}

\put(14.50 ,3.50){\line(1,0){0.70}}
 \put(15.70,3.50){\line(1,0){0.70}}
  \put(17.00,3.50){\line(1,0){0.70}}
\put(18.20 ,3.50){\vector(1,0){2.00}}

\put(11.50,10.00){\makebox(0,0)[cc]{$\varrho_m$}}
\put(11.50,7.00){\makebox(0,0)[cc]{$p$}}
\put(19.00,2.00){\makebox(0,0)[cc]{$\omega$}}

\put(31.00,7.00){\oval( 10,4)}

\put(26.00,7.00){\vector(-1,0){2.00}}
\put(31.00,10.50){\vector(0,-1){1.50}}
\put(31.00,3.50){\vector(0,1){1.50}}

\put(31.00,7.00){\makebox(0,0)[cc]{$\sigma_\perp$}}
\put(25.00,5.00){\makebox(0,0)[cc]{$e$}}

\end{picture}
\caption{Scattering of equivalent photons, forming the ``disk"
with radius $\varrho_m$, on the electron beam with radius
$\sigma_\perp$. }
 \label{fig:3}
\end{figure}

In the reference frame connected with the collider, the equivalent
photon with energy $\hbar \omega$ and the electron with energy
$E_e \gg \hbar \omega$ move toward each other (Fig.~\ref{fig:3})
and perform a Compton scattering. The characteristics of this
process are well known. The main contribution to the Compton
scattering is given by the region where the scattered photons fly
in a direction opposite to that of the initial photons. For such a
backward scattering, the energy of the equivalent photon $\hbar
\omega$ and the energy of the final photon $E_\gamma$ and its
emission angle $\theta_\gamma$ are related by
 \begin{equation}
    \hbar \omega = {E_\gamma \over 4 \gamma^2_e (1 - E_\gamma/E_e )}
   \left[1+ (\gamma_e\theta_\gamma)^2 \right]
 \label{1.1a}
 \end{equation}
and, therefore,
\begin{equation}
 \hbar \omega \sim {E_\gamma \over 4 \gamma^2_e (1 - E_\gamma/E_e
   )}\,.
 \label{1.1}
 \end{equation}

As a result, we find the radius of the ``disk'' of equivalent
photons with the frequency $\omega$ (corresponding to a final
photon with energy $E_\gamma$) as follows:
 \begin{equation}
 \varrho_m = {\gamma_p c\over  \omega} \sim \lambda_e\; {4 \gamma_e
\gamma_p}\, {E_e - E_\gamma \over E_\gamma} \,,\;\;\;
\lambda_e={\hbar\over m_ec}=3.86\cdot 10^{-11}\; {\rm cm}\,.
 \label{size}
 \end{equation}
For the HERA collider with $E_p=820$ GeV and $E_e=28$ GeV one
obtains
 \begin{equation}
 \varrho_m \gtrsim 1\; {\rm cm \ \ \ for\ \ \ } E_\gamma \lesssim
0.2 \;{\rm GeV}\ .
 \label{1.3}
 \end{equation}
Equation (\ref{size}) is also valid for the $e^- e^+ \rightarrow
e^- e^+ \gamma$ process with replacement protons by positrons. For
the VEPP-4 collider it leads to
 \begin{equation}
 \varrho_m \gtrsim 1\; {\rm cm \ \ \ for\ \ \ } E_\gamma \lesssim
15\; \mbox{ MeV }\,,
  \label{1.4}
  \end{equation}
for the PEP-II and KEKB colliders we have
 \begin{equation}
 \varrho_m \gtrsim 1\; {\rm cm \ \ \ for\ \ \ } E_\gamma \lesssim
0.1\; \mbox{ GeV }\,.
  \label{1.4a}
  \end{equation}

The  standard  calculation corresponds to the interaction of the
photons forming the ``disk'' with the unbounded flux of electrons.
However, the particle beams at the HERA collider have finite
transverse beam sizes of the order of $\sigma_\perp\sim 10^{-2}$
cm. Therefore, the equivalent photons from the region
$\sigma_\perp \lesssim \varrho \lesssim \varrho_m$ cannot interact
with the electrons from the other beam. This leads to the
decreasing number of the observed photons and the  ``observed
cross section''  $d \sigma_{\rm obs}$ is smaller than the standard
cross section $d \sigma$ calculated for an infinite transverse
extension of the electron beam,
 \begin{equation}
   d \sigma_{\rm obs} = d \sigma - d \sigma_{\rm cor}.
  \label{1.5}
  \end{equation}
Here the correction $d \sigma_{\rm cor}$ can be presented in the
form
 \begin{equation}
 d \sigma_{\rm cor} = d \sigma_{\rm C}(\omega,\,E_\gamma) \
dn(\omega)
  \label{1.6}
  \end{equation}
where $dn(\omega)$ denotes the number of ``missing'' equivalent
photons and $d \sigma_{\rm C}$ is the cross section of the Compton
scattering.  Let us stress that the equivalent photon
approximation in this region has a high accuracy (the neglected
terms are of the order of $1/\gamma_p$). But for the qualitative
description it is sufficient to use the logarithmic approximation
in which this number is (see\cite{BLP}, \S 99)
 \begin{equation}
  dn = {\alpha \over \pi} {d\omega \over \omega} {d q_{\perp}^2 \over
q_{\perp}^2} \,.
 \label{1.7}
 \end{equation}
Since $q_\perp  \sim 1 / \varrho$, we can present the number of
``missing'' equivalent photons in the form
 \begin{equation}
 dn = {\alpha \over \pi} {d \omega \over \omega} { d\varrho^2 \over
\varrho^2}
 \label{1.8}
 \end{equation}
with the integration region in $\varrho$:
 \begin{equation}
 \sigma_\perp \lesssim \varrho \lesssim \varrho_m = {\gamma_p c
\over \omega}\,.
 \label{1.9}
 \end{equation}
As a result, this number is equal to
 \begin{equation}
 dn(\omega) = 2 {\alpha \over \pi} { d\omega \over \omega} \ln
{\varrho_m \over \sigma_\perp }\,,
 \label{1.10}
 \end{equation}
and the correction to the standard cross section with logarithmic
accuracy is (more exact expression is given by Eq. (\ref{18}))
\footnote{Within this approximation, the standard cross section
has the form (more exact expression is given by Eq. (\ref{17}))
 $$
d\sigma = d \sigma_{\rm C} {\alpha \over \pi} {d\omega \over
\omega} {d q_{\perp}^2 \over q_{\perp}^2}={16\over 3} \alpha
r^2_e\, {dy\over y}\, \left(1-y+\mbox{${3\over 4}$} y^2\right)
\ln{4\gamma_e \gamma_p (1-y)\over y }
 $$
with the integration region $\hbar \omega /(c \gamma_p) \lesssim
\hbar q_\perp \lesssim m_e c$ corresponding to the impact
parameters $\varrho$ in the interval $\lambda_e \lesssim \varrho
\lesssim \varrho_{m}$. }
 \begin{equation}
  d\sigma_{\rm cor} = {16\over 3} \alpha r^2_e\, {dy\over y}\,
\left(1-y+\mbox{${3\over 4}$} y^2\right) \ln{4\gamma_e \gamma_p
(1-y)\lambda_e\over y \sigma_\perp}\,, \;\;y={E_\gamma\over
E_e}\,.
  \label{1.11}
 \end{equation}

\section{MD-effect for PEP-II and KEKB}

Usually in experiments the cross section is found as the ratio of
the number of observed events per second $d\dot N$ to the
luminosity $L$. Also, in our case it is convenient to introduce
the ``observed cross section'', defined as the ratio
 \begin{equation}
d\sigma_{{\rm obs}} ={d\dot N \over L}\,.
 \end{equation}
Contrary to the standard cross section $d\sigma$, the observed
cross section $ d\sigma_{{\rm obs}}$ depends on the parameters of
the beams which scatter. To indicate explicitly this dependence we
introduce the ``correction cross section''  $d\sigma_{{\rm cor}}$
as the difference between $d\sigma$ and $ d\sigma_{{\rm obs}}$:
 \begin{equation}
d\sigma_{{\rm obs}} = d\sigma - d\sigma_{{\rm cor}}\,.
 \end{equation}
The relative magnitude of the MD-effect is given, therefore, by
quantity
 \begin{equation}
 \delta = { d\sigma_{{\rm cor}}\over d\sigma} \,.
 \label{delta}
 \end{equation}
Let us consider the number of photons emitted by electrons in the
process $e^-e^+ \to e^- e^+ \gamma$. The standard cross section
for this process is well known:
 \begin{equation}
d\sigma = {16\over 3} \alpha r^2_e\, {dy\over y}\,
\left(1-y+{3\over 4} y^2\right)\,\left[\ln{4\gamma_e \gamma_p
(1-y)\over y} \, -\, {1\over 2}\right]\,,\;\; y={E_\gamma\over
E_e}\,.
 \label{17}
 \end{equation}
The correction cross section is given by expression
 \begin{equation}
d\sigma_{\rm cor} = {16\over 3} \alpha r^2_e\, {dy\over y}\,
\left[\left(1-y+{3\over 4} y^2\right) L_{\rm cor}
 -{1-y\over 12}\right]
 \label{18}
 \end{equation}
where
 \begin{eqnarray}
L_{\rm cor}&=& \ln{2\sqrt{2}\gamma_e \gamma_p
(1-y)(a_H+a_V)\lambda_e \over  a_H a_V y} \, -\, {3+C\over
 2}\,,
 \nonumber\\
\lambda_e&=&{\hbar\over m_ec}=3.86\cdot 10^{-11}\; {\rm cm}\,,
 \end{eqnarray}
$C=0.577...$, quantities $a_H=\sqrt{\sigma_{eH}^2+ \sigma_{pH}^2}$
and $a_V=\sqrt{\sigma_{eV}^2+ \sigma_{pV}^2}$ related to the
r.m.s. transverse horizontal and vertical bunch sizes
$\sigma_{jH}$ and $\sigma_{jV}$ for the electron, $j=e$, and
positron, $j=p$, beams. In calculations we used data
from Review of Particle Physics--2002~\cite{RPP} (see Table 1).

\centerline{Table 1}
\begin{table}[htb]
\begin{center}
\par
\renewcommand{\arraystretch}{1.5}
\begin{tabular}{|c|c|c|c|c|c|c|c|c|c|} \hline
&$E_e$, & $E_p$, & $\sigma_V$, &  $\sigma_H$, & Energy & $L,
\;10^{33}$ & $N_e,$ & $n_b$ & $\tau_L$,
\\ 
& GeV &  GeV & $\mu$m & $\mu$m & spread, \% & cm$^{-2}\,$s$^{-1}$
& $10^{10}$  & & hr
\\ \hline
PEP-II  & 9 & 3.1 & 4.7 & 157 & 0.061 & 4.6 & 2.1 & 800& 2.5 \\
\hline
 KEKB  & 8 & 3.5 & 2.7 & 110 & 0.07 & 7.25 & 4.5 & 1224 & 3.4 \\ \hline
\end{tabular}
\end{center}
  \label{tablea}
 \end{table}

The observed number of photons is smaller due to MD-effect than
the number of photons calculating without this effect. The
relative magnitude of MD-effect is given by quantity $\delta$ from
Eq. (\ref{delta}) (see Table 2). It is seen that MD-effect reduced
considerable the differential cross section.

\begin{table}[htb]

 \centerline{Table 2}
\vspace{5mm}
\begin{center}
\par
\renewcommand{\arraystretch}{1.5}
\begin{tabular}{|c|c|c|c|c|c|c|} \hline
$y=E_\gamma/E_e$ & 0.001 & 0.005 & 0.01 & 0.05 & 0.1 & 0.5
\\ \hline
$\delta,\; \%$ PEP-II & 31 & 26 & 24 & 19 & 16 & 6.0 \\ \hline
$\delta,\; \%$ KEKB   & 33 & 29 & 26 & 21 & 18 & 8.9 \\ \hline
\end{tabular}
\end{center}
  \label{table2}
 \end{table}

To estimate the integrated contribution of the discussed process
into particle losses, we should integrate the differential
observed cross section from some minimal photon energy. It is
usually assumed that an electron leaves the beam when it emits the
photon with the energy $10$ times larger than the beam energy
spread. In other words, the relative photon energy should be
$y=E_\gamma/E_e \geq y_{\min}$ where $y_{\min}= 0.0061$ for PEP-II
and $y_{\min}= 0.007$ for KEKB. After integration of the observed
cross section from $y_{\min}\ll 1$ up to $y_{\max}=1$, we obtain
\begin{eqnarray}
\sigma_{\rm obs}& =& {16\over 3} \alpha r^2_e\,\left\{
 \left(\ln{1\over y_{\min}}- {5\over 8}\right)
\left[ \ln{\sqrt{2} a_H a_V \over (a_H+a_V)\lambda_e} \, +\,
{2+C\over
 2} \right] \right.
 \nonumber\\
 &+& \left.
 {1\over 12}\left(\ln{1\over y_{\min}}- 1 \right)
 \right\}\,.
 \end{eqnarray}
Let us note that the standard cross section integrated over the
same interval of $y$,
 \begin{eqnarray}
\sigma& =&{16\over 3} \alpha r^2_e\,\left\{\left(\ln{1\over
y_{\min}}- {5\over 8}\right) \,\left[\ln{(4\gamma_e \gamma_p)} \,
-\, {1\over 2}\right]+ {1\over 2}\, \left(\ln{1\over
y_{\min}}\right)^2  \right.
 \nonumber\\
 &-& \left.{3\over 8}-{\pi^2\over 6}\right\}\,,
 \end{eqnarray}
is larger than the observed cross section by about $20$ \% (see
Table 3).

To understand the importance of the bremsstrahlung channel for
particle losses, we estimate the corresponding partial beam
lifetime. The number of particles, which the single electron bunch
losses during a second, equals to
 \begin{equation}
\Delta \dot{N}_e= L\sigma_{\rm obs}/n_b
 \end{equation}
where $L$ is a luminosity and $n_b$ is the number of bunches.
Therefore, the partial lifetime of the electron bunch,
corresponding to bremsstrahlung process at a given luminosity, can
be estimated as
 \begin{equation}
\tau^e_{\rm brem} = {N_e\over \Delta \dot{N}_e} = {N_e n_b\over
L\sigma_{\rm obs}}\,.
 \end{equation}
The obtained numbers for the electron and positron beams are
presented in Table 3. They can be compared with the luminosity
lifetime $\tau_L$ from Table 1 which is some average
characteristics of lifetimes for both beams. More detailed
comparison with the experimental numbers for lifetimes of beams at
KEKB shows that the bremsstrahlung process is important for the
electron beam lifetime, but has rather  small influence on the
positron beam lifetime.

\begin{table}[htb]

 \centerline{Table 3}
\vspace{5mm}
\begin{center}
\par
\renewcommand{\arraystretch}{1.5}
\begin{tabular}{|c|c|c|c|c|c|} \hline
&$\sigma_{\rm obs}$, $10^{-25}$ cm$^2$ & $\sigma/\sigma_{\rm obs}$
& $\tau^e_{\rm brem}$, hr & $\tau^p_{\rm brem}$, hr
\\ \hline
 PEP-II & 2.5 & 1.20 & 4 & 12   \\ \hline
 KEKB  & 2.4 & 1.23 & 8.9 & 14 \\ \hline
\end{tabular}
\end{center}
  \label{table3}
 \end{table}

 \vspace{5mm}

\section*{Acknowledgments}

We are very grateful to A.~Bondar, S.~Heifets, I.~Koop, A.~Onuchin
and E.~Perevedenzev for useful discussions.

\end{document}